\pdfoutput=1

\documentclass[11pt]{article}
\usepackage[utf8]{inputenc}
\usepackage{CJKutf8}
\usepackage{multirow}
\usepackage[final]{acl}

\usepackage{times}
\usepackage{latexsym}
\usepackage{amsmath}
\usepackage[T1]{fontenc}

\usepackage[utf8]{inputenc}

\usepackage{microtype}

\usepackage{inconsolata}

\usepackage{graphicx}
\usepackage{booktabs}
\usepackage[title]{appendix}
\title{Empowering Global Voices: A Data-Efficient, Phoneme-Tone Adaptive Approach to High-Fidelity Speech Synthesis}

\author{%
  Yizhong Geng$^{1,3}$ \
  Jizhuo Xu$^{1,2}$ \
  Zeyu Liang$^{1,3}$ \
  Jinghan Yang$^{1,3}$ \
  Xiaoyi Shi$^{1,4}$ \
  Xiaoyu Shen$^{5}$\thanks{Corresponding can be sent to xyshen@eitech.edu.cn} \\
  $^1$Logic Intelligence Technology\qquad $^2$ Tsinghua University\\
  $^3$Beijing University of Posts and Telecommunications \qquad $^4$Peking University\\
  $^5$Ningbo Key Laboratory of Spatial Intelligence and Digital Derivative, Institute of Digital Twin, EIT\\
}

\begin{document}
\begin{CJK}{UTF8}{gbsn}
\maketitle
\begin{abstract}
Text-to-speech (TTS) technology has achieved impressive results for widely spoken languages, yet many under-resourced languages remain challenged by limited data and linguistic complexities. In this paper, we present a novel methodology that integrates a data-optimized framework with an advanced acoustic model to build high-quality TTS systems for low-resource scenarios. We demonstrate the effectiveness of our approach using Thai as an illustrative case, where intricate phonetic rules and sparse resources are effectively addressed. Our method enables zero-shot voice cloning and improved performance across diverse client applications, ranging from finance to healthcare, education, and law. Extensive evaluations—both subjective and objective—confirm that our model meets state-of-the-art standards, offering a scalable solution for TTS production in data-limited settings, with significant implications for broader industry adoption and multilingual accessibility. All demos are available in \url{https://luoji.cn/static/thai/demo.html}.
\end{abstract}

\section{Introduction}

Recent advancements in text-to-speech (TTS) synthesis have achieved near-human quality for widely spoken languages like English and Mandarin, enabling industrial adoption in customer service, audiobooks, and virtual assistants \cite{anastassiou2024seed}. Yet this progress remains inaccessible to over 7,000 global languages, particularly those with limited labeled speech data~\cite{shen2023xpqa,adelani2024sib}. For linguistically complex languages such as Thai—characterized by tonal distinctions and ambiguous orthography—the scarcity of high-quality training corpora exacerbates the digital divide, stifling equitable access to speech technologies \cite{lux2024meta}.

While LLM-driven TTS systems leverage massive datasets to dynamically adjust pronunciation and prosody \cite{lajszczak2024base}, their data-intensive nature renders them impractical for under-resourced languages \cite{xu2020lrspeech}. To address this gap, we propose a data-efficient framework that combines text-centric training with phoneme-tone adaptive modeling, emulating LLM-level contextual awareness without requiring extensive datasets~\cite{li2023phoneme}. Our approach explicitly targets the dual challenges of low-resource TTS: (1) modeling intricate linguistic features (e.g., tone, phoneme ambiguity) and (2) achieving industrial-grade scalability with minimal data.

\begin{figure*}[htbp]
    \centering
    \includegraphics[width=\textwidth]{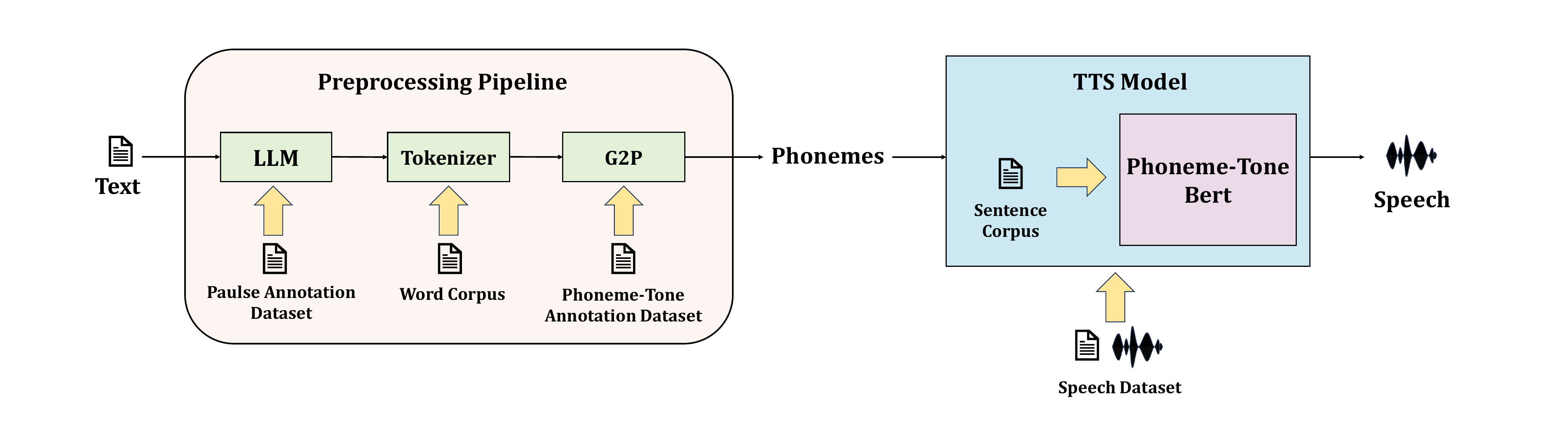}
    \caption{Overview of the Data-Optimized Framework Combined with Advanced Acoustic Model
The architecture comprises two components: (1) the Preprocessing Pipeline (LLM → Tokenizer → grapheme-to-phoneme (G2P)), which converts raw text to phoneme-tone sequences; and (2) the TTS Model, where the Phoneme-Tone Bert module refines contextual pronunciation using text corpus inputs, integrated with acoustic modeling for speech synthesis.}
    \label{fig:arc}
\end{figure*}

Thai, despite being under-resourced, is a language of substantial industrial and demographic importance. It features an intricate five-tone system that requires precise fundamental frequency control—where even minor shifts can alter lexical meaning (e.g., `` Suea '' as `` mat'' [tone 3] versus `` clothes'' [tone 5]~\cite{wutiwiwatchai2017thai})—and grapheme-to-phoneme ambiguities compounded by the absence of clear spoken-word boundaries \cite{christophe2016superseded}. Moreover, Thai is spoken by millions and serves as the official language of a rapidly developing economy with significant regional influence. Its limited speech corpus, orders of magnitude smaller than that of English \cite{thangthai2020tsync}, underscores the urgency of developing efficient TTS frameworks that can unlock considerable industrial value and enhance communication across sectors.

To address this challenge, we have built a comprehensive, multi-dimensional Thai TTS dataset, which forms the foundation for training and validating our TTS system under realistic, industrial-scale conditions. As illustrated in \author{fig:arc}, our system consists of two synergistic components: (1) \textbf{Preprocessing Pipeline}: A robust pipeline that transforms raw Thai text into structured phoneme-tone sequences. This pipeline resolves Thai’s linguistic complexities—including ambiguous word boundaries and intricate tonal patterns—through modules for pause prediction, word segmentation, and grapheme-to-phoneme conversion;
(2) \textbf{TTS Model}: An advanced speech synthesis model that integrates pre-trained audio feature extractors, a GAN-based decoder, and a predictive module for duration, pitch, and energy. The model leverages contextual prosody and style embeddings to dynamically adjust pronunciation and prosody, ensuring high-fidelity synthesis even with limited training data.

Our primary contributions encompass:

\begin{itemize}
\item \textbf{Comprehensive Dataset Construction}: We developed a large-scale, multi-dimensional dataset tailored for Thai TTS, encompassing 500 hours of multi-domain speech, a million-sentence Thai text, and detailed annotations.
\item \textbf{Industry-Usable TTS System}: We deliver the first zero-shot Thai TTS system that achieves state-of-the-art performance, validated through rigorous objective and subjective evaluations across diverse client scenarios (e.g., finance, healthcare, education, law).
\item \textbf{Innovative Technical Strategies}: Our framework leverages a novel data-optimized approach combined with advanced acoustic modeling, including phoneme-tone adaptive modeling. This allows the system to precisely capture Thai’s five-tone system and handle grapheme-to-phoneme ambiguities, all while significantly reducing data demands.
\end{itemize}

\section{Related Work}

\paragraph{TTS: Text to Speech}
Modern TTS technologies, such as FastSpeech2 \cite{ren2020fastspeech} and VITS \cite{kim2021conditional}, have significantly improved speech synthesis in well-resourced languages using sequence-to-sequence architectures and neural vocoders. However, these models struggle with languages like Thai, which have complex tonal systems and preprocessing challenges \cite{thubthong2002tone,shen2017estimation,su2018dialogue}. Their inability to handle tonal variations and limited datasets make them less effective for complex language synthesis \cite{yang2024gigaspeech}.
In contrast, LLM-based models like SeedTTS and CosyVoice \cite{du2024cosyvoice2} offer superior performance but are highly dependent on large-scale datasets for training, making them difficult and costly to deploy for low-resource languages~\cite{su2024unraveling}. The significant data requirements of LLM-driven approaches pose challenges for languages with limited speech data, such as Thai~\cite{xu2020data,zhang2022mdia,zhu2023weaker}.

\paragraph{Thai TTS Challenges}
Thai TTS development faces substantial linguistic and technical hurdles. Unlike English, Thai is a tonal language with five distinct tones, necessitating precise modeling to ensure intelligibility and naturalness \cite{thubthong2002tone,triyason2012perceptual}. Moreover, Thai text lacks explicit word boundaries, complicating word segmentation and pause prediction, which directly impact prosody and fluency \cite{chay2023character}. Existing Thai TTS systems often exhibit incorrect pauses and unnatural intonation due to these ambiguities \cite{wutiwiwatchai2017thai, pipatanakul2024typhoon}, and the limited availability of large, high-quality speech datasets further hinders model training~\cite{shen2022low}. While some Thai TTS approaches rely on rule-based or statistical methods, they fail to fully capture the complexity of Thai phonology and syntax.

\section{Dataset}

This study constructs a comprehensive, multi-dimensional Thai TTS dataset designed to support industrial-scale speech synthesis under low-resource conditions. The dataset is organized into three key categories: Speech Data, Thai Text Data, and Annotation Data. An overview of the datasets is provided in Table~\ref{tab:dataset_overview}.

\paragraph{Speech Dataset}
The Speech Dataset comprises two parts: a multi-domain dataset and a vertical domain dataset. The multi-domain dataset consists of 500 hours of speech from diverse sources. This dataset is designed to enhance the overall TTS capability and zero-shot performance of the model. In addition, the vertical domain dataset includes 40 hours of speech covering specialized fields including finance, healthcare, education, and law, ensuring that the TTS model produces precise pronunciations for domain-specific vocabulary. Detailed production processes and data proportions are provided in Appendix~\ref{sec:Speech Dataset}.

\paragraph{Thai Text Dataset}
The Thai Text Dataset is divided into a sentence corpus and a word corpus. The sentence corpus, containing 1,000,000 sentences, is utilized for training the Phoneme-Tone Bert module to improve contextual prosody modeling. The word corpus, derived from existing lexicons and expanded with manually curated vocabulary, supports the training of the tokenizer, thereby addressing the challenges posed by Thai's unspaced orthography. Detailed information on the curation and processing of the Thai Text Dataset is provided in Appendix~\ref{sec:Thai Text Dataset}.

\paragraph{Annotation Dataset}
The Annotation Dataset provides critical linguistic supervision to resolve Thai-specific synthesis challenges. It includes (1) Pause Annotation, where 15,000 sentences are manually annotated with prosodic boundaries by professional announcers, ensuring accurate pause prediction, and (2) Phoneme-Tone Annotation, comprising 40,000 words, offers detailed IPA phoneme and tone markings to enhance grapheme-to-phoneme conversion and tone modeling. Further details on the annotation procedures and quality control measures are in Appendix~\ref{sec:Annotation Dataset}.

\begin{table}[ht]
  \small
  \centering
  \begin{tabular}{l|l}
    \hline
    \textbf{Dataset} & \textbf{Size} \\
    \hline
    Multi-domain Speech Dataset & 500 hours \\
    Vertical Domain Speech Dataset & 40 hours \\
    Thai Sentence Corpus & 1,000,000 sentences \\
    Thai Word Corpus & 100,000 words \\
    Pause Annotation Dataset & 15,000 sentences \\
    Phoneme-Tone Annotation Dataset & 40,000 words \\
    \hline
  \end{tabular}
  \caption{Overview of the datasets used in this study.}
  \label{tab:dataset_overview}
\end{table}

\section{Preprocessing Pipeline}
The preprocessing stage transforms raw Thai text into annotated phoneme sequences through three sequential modules: 1) a pretrained LLM trained on the Pulse Annotation Dataset to predict prosodic pauses in unpunctuated text, 2) a Tokenizer guided by the Word Corpus to segment unspaced Thai orthography into words, and 3) a G2P converter leveraging the Phoneme-Tone Annotation Dataset to map graphemes to IPA phonemes with tone markers. This pipeline resolves Thai’s linguistic complexities and outputs structured phoneme-tone sequences, enabling robust low-resource TTS.

\paragraph{Pretrained LLM for Pause Prediction}
To address the absence of explicit punctuation and context-dependent pauses in Thai text, we implemented a supervised fine-tuning (SFT) approach using the Pulse Annotation Dataset, a curated corpus of 15,000 Thai sentences annotated with single-type pause positions. The Typhoon2-3B-Instruct \cite{pipatanakul2024typhoon} model was adapted to predict linguistically appropriate pauses by training on instruction-formatted QA pairs. Each training instance included a system prompt ("You are a Thai pause predictor; insert tags <SPACE> based on Thai speech habits").

\paragraph{Tokenizer}
To address Thai’s unspaced orthography and improve segmentation accuracy for domain-specific vocabulary, we extended the pythainlp tokenizer \cite{phatthiyaphaibun2023pythainlp} by augmenting its lexicon from 60,000 to 100,000 words using a word corpus. The expanded vocabulary integrates modern terms through a hybrid approach combining statistical frequency analysis and rule-based morphological patterns. 

\begin{figure*}[htbp]
    \centering
    \includegraphics[width=\textwidth]{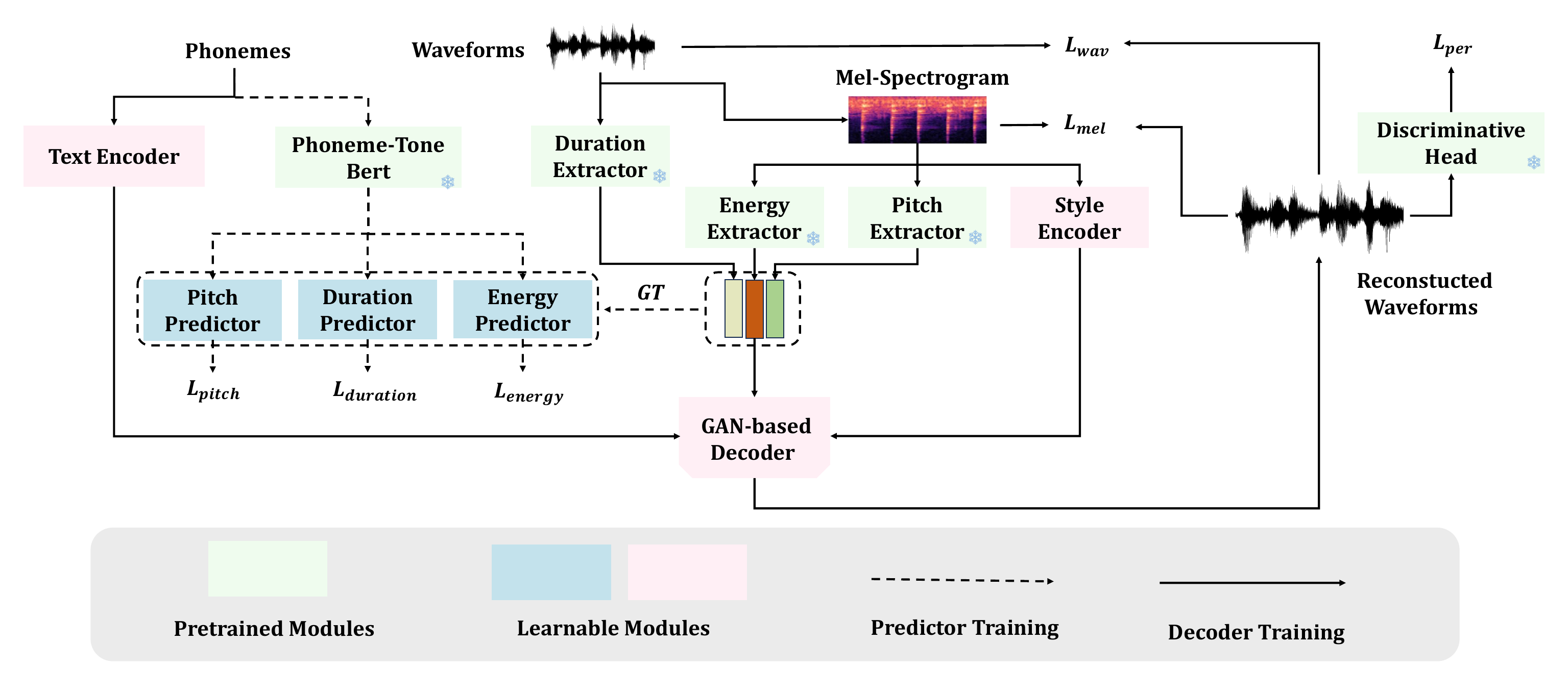}
    \caption{Overview of the proposed TTS model, comprising audio feature extractors, a GAN-based decoder, and a prediction module. The diagram illustrates the different training stages.}
    \label{fig:train}
\end{figure*}

\paragraph{Grapheme-Phoneme Conversion}
To address Thai’s intricate tonal and script complexities, we built a G2P system based on the International Phonetic Alphabet (IPA) \cite{brown2012international}, incorporating Thai’s five-tone markers (mid, low, falling, high, rising). Leveraging the Phoneme-Tone Annotation Dataset—a curated corpus of word-phoneme pairs—we established pronunciation rules covering tone-consonant interactions and contextual exceptions. After tokenization, segmented words are mapped to phonemes via a hybrid approach: rule-based alignment for regular patterns and a transformer model for ambiguous cases.

\section{TTS Model}
Our TTS model (Fig.~\ref{fig:train}) consists of three main components: audio feature extractors, a GAN-based decoder, and a prediction module. The feature extractors, pre-trained on multilingual datasets (e.g., AiShell \cite{fu2021aishell}, LibriSpeech \cite{panayotov2015librispeech}, JVS corpus \cite{takamichi2019jvs}, and KsponSpeech \cite{bang2020ksponspeech}), extract forced alignment, pitch, and energy features from audio/mel-spectrograms. A style encoder embeds audio style into latent vectors. The GAN-based decoder generates waveforms directly from phoneme sequences and the corresponding duration, pitch, energy features, and style vectors, optimizing losses in both time and frequency domains. The prediction module forecasts duration, pitch, and energy from the phoneme sequence. To enhance semantic and prosodic encoding, we label phonemes with tone information per syllable and train a Prosody BERT \cite{devlin2019bert} to encode the phoneme-tone sequence; this representation, combined with the style vector, informs the predictions. After initial separate training, the prediction module is co-trained with the decoder to further improve synthesis quality.

\paragraph{Pretrained Feature Extractor}
We employ three pre-trained models to extract duration, pitch, and energy from waveforms or mel-spectrograms. Given the shared phoneme inventory across languages and the weak correlation between pitch/energy and specific languages, these extraction models are first pre-trained on a multilingual corpus, then fine-tuned on Thai data to address the scarcity of speech resources. Their outputs serve as ground truth to guide predictor training in subsequent stages.

\begin{table*}
  \centering
  \begin{tabular}{llccccc}
    \hline
    \textbf{System} & \textbf{Type} & \textbf{WER (\%)} $\downarrow$ & \textbf{STOI} $\uparrow$ & \textbf{PESQ} $\uparrow$ & \textbf{UTMOS} $\uparrow$ & \textbf{NMOS} $\uparrow$ \\
    \hline
    \textbf{Ours} & Open & \textbf{6.3} \, (\textbf{6.5}) & \textbf{0.92} \, (\textbf{0.94}) & \textbf{4.3} \, (\textbf{4.5}) & \textbf{4.2} \, (\textbf{4.1}) & \textbf{4.4} \, (\textbf{4.6}) \\
    Typhoon2-Audio & Open & 7.8 \, (12.5) & 0.90 \, (0.88) & 4.0 \, (4.0) & 3.5 \, (3.4) & 4.1 \, (4.1) \\
    Seamless-M4T-v2 & Open & 12.3 \, (24.3) & 0.80 \, (0.75) & 3.0 \, (2.8) & 3.0 \, (2.9) & 3.1 \, (3.0) \\
    MMS-TTS & Open & 28.9 \, (35.5) & 0.65 \, (0.60) & 2.5 \, (2.3) & 2.5 \, (2.4) & 2.6 \, (2.5) \\
    PyThaiTTS & Open & 40.3 \, (65.2) & 0.60 \, (0.55) & 2.0 \, (1.8) & 2.0 \, (1.9) & 2.1 \, (2.0) \\
    \hline
    Google TTS & Proprietary & 6.5 \, (14.5) & 0.91 \, (0.85) & 4.1 \, (3.8) & 4.1 \, (3.8) & 4.2 \, (4.0) \\
    Microsoft TTS & Proprietary & 7.1 \, (13.4) & 0.90 \, (0.84) & 4.0 \, (3.7) & 4.0 \, (3.7) & 4.1 \, (3.9) \\
    \hline
  \end{tabular}
  \caption{TTS performance under both general (outside parentheses) and domain-specific (inside parentheses) scenarios. The domain-specific set comprises authentic samples from finance, healthcare, education, and law, reflecting real-world industrial use. Systems labeled as ``Open'' are open-source, while those labeled as ``Proprietary'' are commercial industry standards.}
  \label{tab:compact_results}
\end{table*}

\paragraph{Decoder Training}
To enable cloning capabilities, we introduce a style embedding module that extracts a style vector $s$ from the input waveform. During decoder training, for each audio $w$ and its corresponding text $t$, pre-trained models extract duration $d$, pitch $p$, energy $e$, and obtain phoneme embeddings ($phoneme\_{embed}$) via the text encoder. The waveform decoder $\mathcal{D}$ then reconstructs the waveform as follows:
\[
\hat{w} = \mathcal{D}(phoneme\_{embed}, d, p, e, s)
\]
The reconstruction loss is defined as:
\[
\mathcal{L}_{\text{recon}} = \lambda_1 \mathcal{L}_{\text{time}} + \lambda_2 \mathcal{L}_{\text{freq}} + \lambda_3 \mathcal{L}_{\text{perceptual}}
\]
where $\mathcal{L}_{\text{time}}$ is the L1 loss between the output and target waveforms, $\mathcal{L}_{\text{freq}}$ measures the difference between mel-spectrograms, and $\mathcal{L}_{\text{perceptual}}$ is the GAN-based perceptual loss. These combined losses guide the model towards superior reconstruction performance.

\paragraph{Phoneme-Tone Bert For Predictor Training}
To forecast duration, pitch, and energy from the input phoneme sequence, we first expand the Thai phoneme inventory by integrating tone information via many-to-one tokens. In our revised g2p strategy, tone data is appended to the last phoneme of each syllable, preserving the original token sequence length. We then process a substantial Thai sentence corpus with this g2p method and train a Phoneme-Tone BERT to generate contextual representations ($p\_bert$). Three predictors—duration, pitch, and energy—utilize $p\_bert$ along with a style vector $s$ for their forecasts. Initially, each predictor is trained independently, subsequently, the predictors and decoder are co-trained using a joint loss:
\[
\mathcal{L}_{\text{joint}} = \mathcal{L}_{\text{duration}} + \mathcal{L}_{\text{pitch}} + \mathcal{L}_{\text{energy}} + \mathcal{L}_{\text{decoder}}
\]

\begin{figure*}[ht]
  \centering
  \includegraphics[width=\linewidth]{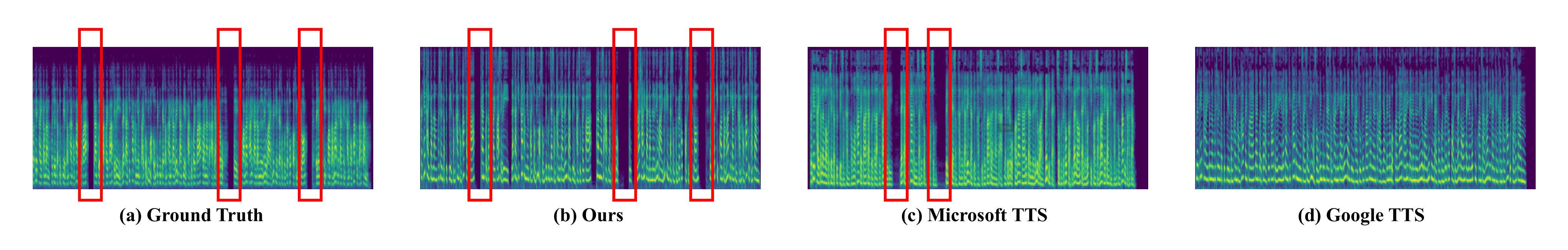}
  \caption{Spectrogram comparison illustrating pause alignment across different TTS systems. The red bounding boxes highlight detected pause regions.}
  \label{fig:pause}
\end{figure*}

\section{Experiments}

\paragraph{Implementation Details}
The pretrained LLM for pause prediction was trained on the Pulse Annotation Dataset, which comprises 15,000 Thai sentences annotated with single-type pause positions. The input sequences were tokenized with a maximum length of 512 tokens. For optimization, we used the AdamW optimizer with coefficients β₁ = 0.9 and β₂ = 0.98, a learning rate of 1e-5, and a weight decay of 0.01. The model converged within approximately 200k training steps using a batch size equivalent to processing 16 sentences per step.

The Phoneme-Tone Bert module was trained on a sentence corpus of 1 million sentences using a 12-layer BERT architecture with 768 hidden units and 12 self-attention heads. We used a masked language modeling objective, AdamW optimizer (learning rate 2e-5, weight decay 0.01), batch size 32, maximum sequence length 256, dropout rate 0.1, and trained for 500k steps.

The TTS Model is trained using the entire speech dataset, which includes 500 hours of multi-domain data and 40 hours of vertical domain data. The training employs the AdamW optimizer with β₁ = 0.9 and β₂ = 0.96. The model undergoes training for 8 days on 8 A800 GPUs, using a batch size of 768 samples.

\paragraph{Effect of Preprocessing Pipeline Modules}

To evaluate each module's contribution, we performed an ablation study by removing them one at a time. Table~\ref{tab:ablation} compares our full model with three variants: (i) no pause optimization, (ii) no tokenization optimization, and (iii) no G2P optimization. We used Word Error Rate (WER) and Naturalness Mean Opinion Score (NMOS) as metrics.

Table~\ref{tab:ablation} shows that pause optimization is crucial for natural prosody, as removing it raises WER from 6.3\% to 6.5\% and lowers NMOS from 4.4 to 3.8. Without tokenization optimization, WER jumps to 10.2\% and NMOS drops to 3.9, highlighting its role in text segmentation. G2P optimization has the greatest impact, with WER at 22.5\% and NMOS at 3.0, indicating poor performance overall. Figure~\ref{fig:pause} provides a spectrogram comparison of different TTS outputs. It illustrates how accurate pause prediction yields better alignment with ground-truth prosody, resulting in clearer and more natural synthesized speech.

\begin{table}
  \small
  \centering
  \begin{tabular}{lcc}
    \hline
    \textbf{System} & \textbf{WER (\%)} $\downarrow$ & \textbf{NMOS} $\uparrow$ \\
    \hline
    \textbf{Ours} & \textbf{6.3} & \textbf{4.4} \\
    w/o Pause Optimization & 6.5 & 3.8 \\
    w/o Tokenization Optimization & 10.2 & 3.9 \\
    w/o G2P Optimization & 22.5 & 3.0 \\
    \hline
  \end{tabular}
  \caption{Ablation study on the preprocessing pipeline. Removing each module reveals its contribution.}
  \label{tab:ablation}
\end{table}

\paragraph{TTS Performance}

Table~\ref{tab:compact_results} summarizes TTS performance on both a general-domain test set and domain-specific samples. The general-domain set is drawn from TSync2, an open-source Thai corpus widely used for benchmarking. For the domain-specific evaluation, we deployed our TTS system in four real-world business scenarios: automated transaction summaries in finance, telehealth voice guidance in healthcare, online course narration in education, and legal document reading in law. End users in each domain rated the synthesized sentences on intelligibility and term accuracy, with their feedback contributing to the NMOS scores reported. This practical assessment highlights our system’s ability to deliver clear, domain-appropriate speech in genuine industry contexts.

Our model achieves the highest overall accuracy and speech quality among open-source systems, showing notable robustness in real-world industrial settings. In contrast, proprietary solutions like Google TTS and Microsoft TTS, while performing competitively on the TSync2 set (WER of 6.5\% and 7.1\%, respectively), exhibit larger performance drops in specialized domains (WER of 14.5\% and 13.4\%). Field professionals also reported higher mispronunciation rates in these proprietary systems, especially for domain-specific jargon. This suggests our approach excels in broad usage scenarios and maintains reliability in high-stakes, industry-specific environments.

\begin{figure}[ht]
  \centering
  \includegraphics[width=\linewidth]{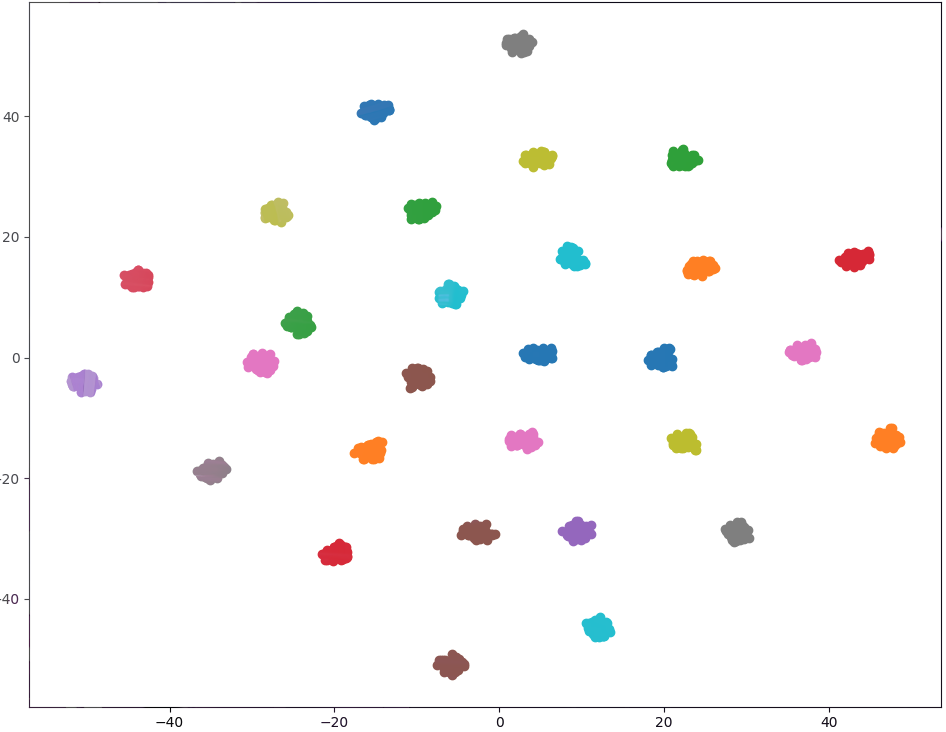}
\caption{t-SNE visualization of speaker embeddings extracted from the synthesized speech. Each point represents a speaker embedding, and distinct clusters show that our zero-shot TTS preserves speaker identity.}
  \label{fig:tsne}
\end{figure}

\paragraph{Zero-shot TTS Performance}

Zero-shot TTS extends conventional TTS by synthesizing speech for previously unseen speakers without additional speaker-specific data or fine-tuning. In other words, it can clone a speaker’s timbre from reference audio, enabling rapid deployment for new voices. Since all baseline models lack this capability, we compare our system with OpenVoice—a widely used voice conversion model \cite{qin2023openvoice}. As shown in Table~\ref{tab:clone_results}, our system attains a SIM of 0.91 and SMOS of 4.5, surpassing OpenVoice’s 0.85 and 4.0. Figure~\ref{fig:tsne} further illustrates this advantage: distinct clusters in the speaker embedding space confirm robust identity preservation without speaker-specific training.

\begin{table}[ht]
  \centering
  \begin{tabular}{lcc}
    \hline
    \textbf{System} & \textbf{SIM} $\uparrow$ & \textbf{SMOS} $\uparrow$ \\
    \hline
    \textbf{Ours} & \textbf{0.91} & \textbf{4.5} \\
    OpenVoice (10s) & 0.85 & 4.0 \\
    \hline
  \end{tabular}
  \caption{Zero-shot TTS performance comparison. SIM (machine acoustic similarity) and SMOS (human-judged speaker identity) highlight our advantage.}
  \label{tab:clone_results}
\end{table}

\section{Conclusion}

We present a data-optimized framework with an advanced acoustic model for TTS in under-resourced languages, using Thai as a representative case. Our pipeline integrates sophisticated preprocessing with a robust TTS model, achieving state-of-the-art results in both general and domain-specific tasks, validated in commercial scenarios across finance, healthcare, education, and law. Experiments confirm notable quality gains and successful zero-shot voice cloning, demonstrating efficacy and business viability. Beyond bridging performance gaps in low-resource contexts, our approach offers a scalable solution adaptable to other under-resourced languages. Future work will extend this framework to other languages with similar constraints.

\bibliography{custom}

\begin{thebibliography}{40}
\providecommand{\natexlab}[1]{#1}

\bibitem[{Adelani et~al.(2024)Adelani, Liu, Shen, Vassilyev, Alabi, Mao, Gao, and Lee}]{adelani2024sib}
David Adelani, Hannah Liu, Xiaoyu Shen, Nikita Vassilyev, Jesujoba Alabi, Yanke Mao, Haonan Gao, and En-Shiun Lee. 2024.
\newblock Sib-200: A simple, inclusive, and big evaluation dataset for topic classification in 200+ languages and dialects.
\newblock In \emph{Proceedings of the 18th Conference of the European Chapter of the Association for Computational Linguistics (Volume 1: Long Papers)}, pages 226--245.

\bibitem[{Anastassiou et~al.(2024)Anastassiou, Chen, Chen, Chen, Chen, Chen, Cong, Deng, Ding, Gao et~al.}]{anastassiou2024seed}
Philip Anastassiou, Jiawei Chen, Jitong Chen, Yuanzhe Chen, Zhuo Chen, Ziyi Chen, Jian Cong, Lelai Deng, Chuang Ding, Lu~Gao, et~al. 2024.
\newblock Seed-tts: A family of high-quality versatile speech generation models.
\newblock \emph{arXiv preprint arXiv:2406.02430}.

\bibitem[{Bang et~al.(2020)Bang, Yun, Kim, Choi, Lee, Kim, Kim, Park, Lee, and Kim}]{bang2020ksponspeech}
Jeong-Uk Bang, Seung Yun, Seung-Hi Kim, Mu-Yeol Choi, Min-Kyu Lee, Yeo-Jeong Kim, Dong-Hyun Kim, Jun Park, Young-Jik Lee, and Sang-Hun Kim. 2020.
\newblock Ksponspeech: Korean spontaneous speech corpus for automatic speech recognition.
\newblock \emph{Applied Sciences}, 10(19):6936.

\bibitem[{Barrault et~al.(2023)Barrault, Chung, Meglioli, Dale, Dong, Duquenne, Elsahar, Gong, Heffernan, Hoffman et~al.}]{barrault2023seamlessm4t}
Lo{\"\i}c Barrault, Yu-An Chung, Mariano~Cora Meglioli, David Dale, Ning Dong, Paul-Ambroise Duquenne, Hady Elsahar, Hongyu Gong, Kevin Heffernan, John Hoffman, et~al. 2023.
\newblock Seamlessm4t: Massively multilingual \& multimodal machine translation.
\newblock \emph{arXiv preprint arXiv:2308.11596}.

\bibitem[{Brown(2012)}]{brown2012international}
Adam Brown. 2012.
\newblock International phonetic alphabet.
\newblock \emph{The encyclopedia of applied linguistics}.

\bibitem[{Chay-intr et~al.(2023)Chay-intr, Kamigaito, and Okumura}]{chay2023character}
Thodsaporn Chay-intr, Hidetaka Kamigaito, and Manabu Okumura. 2023.
\newblock Character-based thai word segmentation with multiple attentions.
\newblock \emph{Journal of Natural Language Processing}, 30(2):372--400.

\bibitem[{Christophe et~al.(2016)Christophe, Junichi, Kirsten et~al.}]{christophe2016superseded}
YJMK~Veaux Christophe, Yamagishi Junichi, MacDonald Kirsten, et~al. 2016.
\newblock Superseded-cstr vctk corpus: English multi-speaker corpus for cstr voice cloning toolkit.

\bibitem[{D{\'e}fossez(2021)}]{defossez2021hybrid}
Alexandre D{\'e}fossez. 2021.
\newblock Hybrid spectrogram and waveform source separation.
\newblock \emph{arXiv preprint arXiv:2111.03600}.

\bibitem[{Devlin et~al.(2019)Devlin, Chang, Lee, and Toutanova}]{devlin2019bert}
Jacob Devlin, Ming-Wei Chang, Kenton Lee, and Kristina Toutanova. 2019.
\newblock Bert: Pre-training of deep bidirectional transformers for language understanding.
\newblock In \emph{Proceedings of the 2019 conference of the North American chapter of the association for computational linguistics: human language technologies, volume 1 (long and short papers)}, pages 4171--4186.

\bibitem[{Doumanidis et~al.(2021)Doumanidis, Anagnostou, Arvaniti, and Papadopoulou}]{doumanidis2021rnnoise}
Constantine~C Doumanidis, Christina Anagnostou, Evangelia-Sofia Arvaniti, and Anthi Papadopoulou. 2021.
\newblock Rnnoise-ex: Hybrid speech enhancement system based on rnn and spectral features.
\newblock \emph{arXiv preprint arXiv:2105.11813}.

\bibitem[{Du et~al.(2024)Du, Wang, Chen, Shi, Lv, Zhao, Gao, Yang, Gao, Wang et~al.}]{du2024cosyvoice2}
Zhihao Du, Yuxuan Wang, Qian Chen, Xian Shi, Xiang Lv, Tianyu Zhao, Zhifu Gao, Yexin Yang, Changfeng Gao, Hui Wang, et~al. 2024.
\newblock Cosyvoice 2: Scalable streaming speech synthesis with large language models.
\newblock \emph{arXiv preprint arXiv:2412.10117}.

\bibitem[{Fu et~al.(2021)Fu, Cheng, Lv, Jv, Kong, Chen, Hu, Xie, Wu, Bu et~al.}]{fu2021aishell}
Yihui Fu, Luyao Cheng, Shubo Lv, Yukai Jv, Yuxiang Kong, Zhuo Chen, Yanxin Hu, Lei Xie, Jian Wu, Hui Bu, et~al. 2021.
\newblock Aishell-4: An open source dataset for speech enhancement, separation, recognition and speaker diarization in conference scenario.
\newblock \emph{arXiv preprint arXiv:2104.03603}.

\bibitem[{Kim et~al.(2021)Kim, Kong, and Son}]{kim2021conditional}
Jaehyeon Kim, Jungil Kong, and Juhee Son. 2021.
\newblock Conditional variational autoencoder with adversarial learning for end-to-end text-to-speech.
\newblock In \emph{International Conference on Machine Learning}, pages 5530--5540. PMLR.

\bibitem[{{\L}ajszczak et~al.(2024){\L}ajszczak, C{\'a}mbara, Li, Beyhan, Van~Korlaar, Yang, Joly, Mart{\'\i}n-Cortinas, Abbas, Michalski et~al.}]{lajszczak2024base}
Mateusz {\L}ajszczak, Guillermo C{\'a}mbara, Yang Li, Fatih Beyhan, Arent Van~Korlaar, Fan Yang, Arnaud Joly, {\'A}lvaro Mart{\'\i}n-Cortinas, Ammar Abbas, Adam Michalski, et~al. 2024.
\newblock Base tts: Lessons from building a billion-parameter text-to-speech model on 100k hours of data.
\newblock \emph{arXiv preprint arXiv:2402.08093}.

\bibitem[{Li et~al.(2023)Li, Han, Jiang, and Mesgarani}]{li2023phoneme}
Yinghao~Aaron Li, Cong Han, Xilin Jiang, and Nima Mesgarani. 2023.
\newblock Phoneme-level bert for enhanced prosody of text-to-speech with grapheme predictions.
\newblock In \emph{ICASSP 2023-2023 IEEE International Conference on Acoustics, Speech and Signal Processing (ICASSP)}, pages 1--5. IEEE.

\bibitem[{Lowphansirikul et~al.(2021)Lowphansirikul, Polpanumas, Jantrakulchai, and Nutanong}]{lowphansirikul2021wangchanberta}
Lalita Lowphansirikul, Charin Polpanumas, Nawat Jantrakulchai, and Sarana Nutanong. 2021.
\newblock Wangchanberta: Pretraining transformer-based thai language models.
\newblock \emph{arXiv preprint arXiv:2101.09635}.

\bibitem[{Lux et~al.(2024)Lux, Meyer, Behringer, Zalkow, Do, Coler, Habets, and Vu}]{lux2024meta}
Florian Lux, Sarina Meyer, Lyonel Behringer, Frank Zalkow, Phat Do, Matt Coler, Emanu{\"e}l~AP Habets, and Ngoc~Thang Vu. 2024.
\newblock Meta learning text-to-speech synthesis in over 7000 languages.
\newblock \emph{arXiv preprint arXiv:2406.06403}.

\bibitem[{Panayotov et~al.(2015)Panayotov, Chen, Povey, and Khudanpur}]{panayotov2015librispeech}
Vassil Panayotov, Guoguo Chen, Daniel Povey, and Sanjeev Khudanpur. 2015.
\newblock Librispeech: an asr corpus based on public domain audio books.
\newblock In \emph{2015 IEEE international conference on acoustics, speech and signal processing (ICASSP)}, pages 5206--5210. IEEE.

\bibitem[{Phatthiyaphaibun et~al.(2023)Phatthiyaphaibun, Chaovavanich, Polpanumas, Suriyawongkul, Lowphansirikul, Chormai, Limkonchotiwat, Suntorntip, and Udomcharoenchaikit}]{phatthiyaphaibun2023pythainlp}
Wannaphong Phatthiyaphaibun, Korakot Chaovavanich, Charin Polpanumas, Arthit Suriyawongkul, Lalita Lowphansirikul, Pattarawat Chormai, Peerat Limkonchotiwat, Thanathip Suntorntip, and Can Udomcharoenchaikit. 2023.
\newblock Pythainlp: Thai natural language processing in python.
\newblock \emph{arXiv preprint arXiv:2312.04649}.

\bibitem[{Pipatanakul et~al.(2024)Pipatanakul, Manakul, Nitarach, Sirichotedumrong, Nonesung, Jaknamon, Pengpun, Taveekitworachai, Na-Thalang, Sripaisarnmongkol et~al.}]{pipatanakul2024typhoon}
Kunat Pipatanakul, Potsawee Manakul, Natapong Nitarach, Warit Sirichotedumrong, Surapon Nonesung, Teetouch Jaknamon, Parinthapat Pengpun, Pittawat Taveekitworachai, Adisai Na-Thalang, Sittipong Sripaisarnmongkol, et~al. 2024.
\newblock Typhoon 2: A family of open text and multimodal thai large language models.
\newblock \emph{arXiv preprint arXiv:2412.13702}.

\bibitem[{Pratap et~al.(2024)Pratap, Tjandra, Shi, Tomasello, Babu, Kundu, Elkahky, Ni, Vyas, Fazel-Zarandi et~al.}]{pratap2024scaling}
Vineel Pratap, Andros Tjandra, Bowen Shi, Paden Tomasello, Arun Babu, Sayani Kundu, Ali Elkahky, Zhaoheng Ni, Apoorv Vyas, Maryam Fazel-Zarandi, et~al. 2024.
\newblock Scaling speech technology to 1,000+ languages.
\newblock \emph{Journal of Machine Learning Research}, 25(97):1--52.

\bibitem[{Qin et~al.(2023)Qin, Zhao, Yu, and Sun}]{qin2023openvoice}
Zengyi Qin, Wenliang Zhao, Xumin Yu, and Xin Sun. 2023.
\newblock Openvoice: Versatile instant voice cloning.
\newblock \emph{arXiv preprint arXiv:2312.01479}.

\bibitem[{Radford et~al.(2023)Radford, Kim, Xu, Brockman, McLeavey, and Sutskever}]{radford2023robust}
Alec Radford, Jong~Wook Kim, Tao Xu, Greg Brockman, Christine McLeavey, and Ilya Sutskever. 2023.
\newblock Robust speech recognition via large-scale weak supervision.
\newblock In \emph{International conference on machine learning}, pages 28492--28518. PMLR.

\bibitem[{Ren et~al.(2020)Ren, Hu, Tan, Qin, Zhao, Zhao, and Liu}]{ren2020fastspeech}
Yi~Ren, Chenxu Hu, Xu~Tan, Tao Qin, Sheng Zhao, Zhou Zhao, and Tie-Yan Liu. 2020.
\newblock Fastspeech 2: Fast and high-quality end-to-end text to speech.
\newblock \emph{arXiv preprint arXiv:2006.04558}.

\bibitem[{Shen et~al.(2023)Shen, Asai, Byrne, and De~Gispert}]{shen2023xpqa}
Xiaoyu Shen, Akari Asai, Bill Byrne, and Adria De~Gispert. 2023.
\newblock xpqa: Cross-lingual product question answering in 12 languages.
\newblock In \emph{Proceedings of the 61st Annual Meeting of the Association for Computational Linguistics (Volume 5: Industry Track)}, pages 103--115.

\bibitem[{Shen et~al.(2017)Shen, Oualil, Greenberg, Singh, and Klakow}]{shen2017estimation}
Xiaoyu Shen, Youssef Oualil, Clayton Greenberg, Mittul Singh, and Dietrich Klakow. 2017.
\newblock Estimation of gap between current language models and human performance.

\bibitem[{Shen et~al.(2022)Shen, Vakulenko, Del~Tredici, Barlacchi, Byrne, and de~Gispert}]{shen2022low}
Xiaoyu Shen, Svitlana Vakulenko, Marco Del~Tredici, Gianni Barlacchi, Bill Byrne, and Adri{\`a} de~Gispert. 2022.
\newblock Low-resource dense retrieval for open-domain question answering: A comprehensive survey.
\newblock \emph{arXiv preprint arXiv:2208.03197}.

\bibitem[{Smith(2007)}]{smith2007overview}
Ray Smith. 2007.
\newblock An overview of the tesseract ocr engine.
\newblock In \emph{Ninth international conference on document analysis and recognition (ICDAR 2007)}, volume~2, pages 629--633. IEEE.

\bibitem[{Su et~al.(2018)Su, Shen, Hu, Li, and Chen}]{su2018dialogue}
Hui Su, Xiaoyu Shen, Pengwei Hu, Wenjie Li, and Yun Chen. 2018.
\newblock Dialogue generation with gan.
\newblock In \emph{Proceedings of the AAAI Conference on Artificial Intelligence}, volume~32.

\bibitem[{Su et~al.(2024)Su, Tian, Shen, and Cai}]{su2024unraveling}
Hui Su, Zhi Tian, Xiaoyu Shen, and Xunliang Cai. 2024.
\newblock Unraveling the mystery of scaling laws: Part i.
\newblock \emph{arXiv preprint arXiv:2403.06563}.

\bibitem[{Takamichi et~al.(2019)Takamichi, Mitsui, Saito, Koriyama, Tanji, and Saruwatari}]{takamichi2019jvs}
Shinnosuke Takamichi, Kentaro Mitsui, Yuki Saito, Tomoki Koriyama, Naoko Tanji, and Hiroshi Saruwatari. 2019.
\newblock Jvs corpus: free japanese multi-speaker voice corpus.
\newblock \emph{arXiv preprint arXiv:1908.06248}.

\bibitem[{Thangthai et~al.(2020)Thangthai, Thatphithakkul, Thangthai, and Namsanit}]{thangthai2020tsync}
Ausdang Thangthai, Sumonmas Thatphithakkul, Kwanchiva Thangthai, and Arnon Namsanit. 2020.
\newblock Tsync-3miti: Audiovisual speech synthesis database from found data.
\newblock In \emph{2020 23rd Conference of the Oriental COCOSDA International Committee for the Co-ordination and Standardisation of Speech Databases and Assessment Techniques (O-COCOSDA)}, pages 77--82. IEEE.

\bibitem[{Thubthong et~al.(2002)Thubthong, Kijsirikul, and Luksaneeyanawin}]{thubthong2002tone}
Nuttakorn Thubthong, Boonserm Kijsirikul, and Sudaporn Luksaneeyanawin. 2002.
\newblock Tone recognition in thai continuous speech based on coarticulaion, intonation and stress effects.
\newblock In \emph{INTERSPEECH}, pages 1169--1172.

\bibitem[{Triyason and Kanthamanon(2012)}]{triyason2012perceptual}
Tuul Triyason and Prasert Kanthamanon. 2012.
\newblock Perceptual evaluation of speech quality measurement on speex codec voip with tonal language thai.
\newblock In \emph{International Conference on Advances in Information Technology}, pages 181--190. Springer.

\bibitem[{Wutiwiwatchai et~al.(2017)Wutiwiwatchai, Hansakunbuntheung, Rugchatjaroen, Saychum, Kasuriya, and Chootrakool}]{wutiwiwatchai2017thai}
Chai Wutiwiwatchai, Chatchawarn Hansakunbuntheung, Anocha Rugchatjaroen, Sittipong Saychum, Sawit Kasuriya, and Patcharika Chootrakool. 2017.
\newblock Thai text-to-speech synthesis: a review.
\newblock \emph{Journal of Intelligent Informatics and Smart Technology}.

\bibitem[{Xu et~al.(2020{\natexlab{a}})Xu, Qiu, Zhang, Wang, Shen, and De~Melo}]{xu2020data}
Binxia Xu, Siyuan Qiu, Jie Zhang, Yafang Wang, Xiaoyu Shen, and Gerard De~Melo. 2020{\natexlab{a}}.
\newblock Data augmentation for multiclass utterance classification--a systematic study.
\newblock In \emph{Proceedings of the 28th international conference on computational linguistics}, pages 5494--5506.

\bibitem[{Xu et~al.(2020{\natexlab{b}})Xu, Tan, Ren, Qin, Li, Zhao, and Liu}]{xu2020lrspeech}
Jin Xu, Xu~Tan, Yi~Ren, Tao Qin, Jian Li, Sheng Zhao, and Tie-Yan Liu. 2020{\natexlab{b}}.
\newblock Lrspeech: Extremely low-resource speech synthesis and recognition.
\newblock In \emph{Proceedings of the 26th ACM SIGKDD International Conference on Knowledge Discovery \& Data Mining}, pages 2802--2812.

\bibitem[{Yang et~al.(2024)Yang, Song, Zhuo, Cui, Li, Yang, Du, Ma, Liu, Wang et~al.}]{yang2024gigaspeech}
Yifan Yang, Zheshu Song, Jianheng Zhuo, Mingyu Cui, Jinpeng Li, Bo~Yang, Yexing Du, Ziyang Ma, Xunying Liu, Ziyuan Wang, et~al. 2024.
\newblock Gigaspeech 2: An evolving, large-scale and multi-domain asr corpus for low-resource languages with automated crawling, transcription and refinement.
\newblock \emph{arXiv preprint arXiv:2406.11546}.

\bibitem[{Zhang et~al.(2022)Zhang, Shen, Chang, Ge, and Chen}]{zhang2022mdia}
Qingyu Zhang, Xiaoyu Shen, Ernie Chang, Jidong Ge, and Pengke Chen. 2022.
\newblock Mdia: A benchmark for multilingual dialogue generation in 46 languages.
\newblock \emph{arXiv preprint arXiv:2208.13078}.

\bibitem[{Zhu et~al.(2023)Zhu, Shen, Mosbach, Stephan, and Klakow}]{zhu2023weaker}
Dawei Zhu, Xiaoyu Shen, Marius Mosbach, Andreas Stephan, and Dietrich Klakow. 2023.
\newblock Weaker than you think: A critical look at weakly supervised learning.
\newblock In \emph{Proceedings of the 61st Annual Meeting of the Association for Computational Linguistics (Volume 1: Long Papers)}, pages 14229--14253.

\end{thebibliography}
\clearpage
\begin{appendices}

\section{Evaluation Metrics}\label{sec:metrics}
This study uses seven principal metrics across four dimensions—accuracy, voice cloning, naturalness, and speech quality/intelligibility—to evaluate system performance.

\textbf{Accuracy} is measured by Word Error Rate (WER), which quantifies transcription fidelity by comparing discrepancies between synthesized speech and reference texts, with lower WER indicating better accuracy.

\textbf{Voice Cloning} is assessed using the Similarity Score (SIM) and Subjective Similarity Mean Opinion Score (SMOS). SIM calculates acoustic similarity using cosine analysis of phonetic-tonal features, while SMOS is based on ratings from fifty native Thai speakers evaluating thirty samples on a 5-point scale.

\textbf{Naturalness} is evaluated with three metrics: the UTokyo-SaruLab Mean Opinion Score (UTMOS), Perceptual Evaluation of Speech Quality (PESQ), and Naturalness Mean Opinion Score (NMOS). UTMOS predicts naturalness by analyzing prosody, spectral stability, and artifacts. PESQ quantifies quality degradation and spectral distortions, while NMOS is based on subjective ratings assessing fluency and prosody from fifty listeners.

\textbf{Speech Intelligibility} is measured by the Short-Time Objective Intelligibility (STOI), which correlates with word recognition rates by analyzing temporal-spectral envelope similarities between synthesized and reference speech, critical for evaluating tone preservation.

\section{Baseline Systems}\label{sec:baseline}
To benchmark the performance of our model, we compare it against multiple baseline systems spanning open-source and proprietary paradigms. The baselines are described below:
\begin{itemize}
  \item \textbf{PyThaiTTS} \cite{phatthiyaphaibun2023pythainlp}: A Thai-optimized Tacotron2 model trained on TSync datasets.
  \item \textbf{Seamless-M4T-v2} \cite{barrault2023seamlessm4t}: A multilingual system supporting Thai among 100+ languages.
  \item \textbf{MMS-TTS} \cite{pratap2024scaling}: A model covering Thai within its 1,100+ language inventory.
  \item \textbf{Typhoon2-Audio} \cite{pipatanakul2024typhoon}: An end-to-end multimodal model that enables parallel speech-text generation through integrated speech encoders and non-autoregressive decoders.
  \item \textbf{Google Cloud TTS (th-TH-Standard-A)}\footnote{\url{https://cloud.google.com/text-to-speech}}: A proprietary, industry-standard commercial solution optimized for Thai TTS.
  \item \textbf{Microsoft Azure TTS (Premwadee)}\footnote{\url{https://azure.microsoft.com/en-us/services/cognitive-services/text-to-speech/}}: A proprietary system offering state-of-the-art Thai TTS performance.
\end{itemize}

\section{Dataset}
\subsection{Speech Dataset}\label{sec:Speech Dataset}

This section details the construction of our Speech Dataset, outlining both the data composition and the processing workflow. The dataset is meticulously curated to ensure industrial-grade quality and linguistic diversity, which are crucial for training robust TTS models.

\subsubsection{Data Composition and Distribution}

\textbf{Multi-domain Corpus:}  
The multi-domain speech data is systematically collected from multiple public resources, ensuring a balanced mix of content and speaker diversity. The dataset comprises four primary data sources:
\begin{itemize}
  \item \textbf{News Broadcasts (30\%)}: Sourced from the Thai Broadcasting Radio \footnote{Source:\url{https://www.radio-thai.com/ }}.
  \item \textbf{Audiobooks (10\%)}: Obtained from open-source speech libraries \footnote{Source:\url{https://www.storytel.com/th/audiobooks}} \footnote{Source:\url{https://www.ookbee.com/shop/audios}}. 
  \item \textbf{Social Media Short Videos (25\%)}: Scraped from TikTok’s public content via compliant APIs.
  \item \textbf{Daily Conversation Podcasts (35\%)}: Crawled from public YouTube channels.
\end{itemize}
The audio adheres to an industrial-grade recording standard with a 24kHz sampling rate and a signal-to-noise ratio (SNR) of at least 35dB. The data includes over 600 speakers, maintains a near-balanced gender ratio of 1.2:1.
Table~\ref{tab:speech_data_composition} provides an overview of the multi-domain data composition (totaling 500 hours).

\begin{table*}[ht]
  \centering
  \small
  \begin{tabular}{l|c|l}
    \hline
    \textbf{Data Source} & \textbf{Percentage} & \textbf{Description} \\
    \hline
    News Broadcasts & 30\% & Thai National Broadcasting Radio \\
    Audiobooks & 10\% & Open-source speech libraries \\
    Social Media Short Videos & 25\% & TikTok public content \\
    Daily Conversation Podcasts & 35\% & Public YouTube channels \\
    \hline
    \multicolumn{3}{l}{\textit{Total: 100\% (500 hours)}} \\
    \hline
  \end{tabular}
  \caption{Data composition of the multi-domain Speech Dataset.}
  \label{tab:speech_data_composition}
\end{table*}

\textbf{Vertical Domain Corpus:}  
In addition to the multi-domain corpus, the Speech Dataset includes a vertical domain corpus consisting of 40 hours of speech data from YouTube open-source content. This subset is specifically collected to capture the nuances of specialized fields and ensure the TTS model produces precise pronunciations for domain-specific vocabulary. The vertical domain data is evenly distributed across four specialized sectors:
\begin{itemize}
  \item \textbf{Finance (25\%)}: Recorded from corporate earnings calls, investor presentations, and financial news.
  \item \textbf{Healthcare (25\%)}: Sourced from medical lectures, healthcare communications, and hospital announcements.
  \item \textbf{Education (25\%)}: Collected from university lectures, academic seminars, and educational podcasts.
  \item \textbf{Law (25\%)}: Derived from court proceedings, legal seminars, and formal legal communications.
\end{itemize}
All vertical domain recordings meet the same industrial-grade standards as the multi-domain data, with a 24kHz sampling rate and a minimum SNR of 35dB.

\subsubsection{Data Processing Workflow}
The raw audio data undergoes a multi-stage processing pipeline to ensure high-quality, clean speech suitable for TTS training:
\begin{enumerate}
  \item \textbf{Noise Separation and Reduction}: Background noise, including music and environmental sounds, is first separated using Demucs v4 \cite{defossez2021hybrid}, followed by residual noise reduction via RNNoise \cite{doumanidis2021rnnoise}.
  \item \textbf{Speech Activity Detection (VAD)}: WebRTC-based VAD \footnote{Source:\url{https://webrtc.org/}}  is employed to segment the audio into clean clips ranging from 5 to 15 seconds.
  \item \textbf{Text Extraction and Verification}: For audio segments lacking corresponding text, hardcoded subtitles are extracted using Tesseract OCR \cite{smith2007overview} and then cross-checked with outputs from Whisper-large-v3 ASR \cite{radford2023robust}. Segments with a character error rate (CER) above 5\% are manually verified.
\end{enumerate}
This comprehensive processing workflow ensures that both the multi-domain and vertical domain corpora are of high quality, facilitating robust and accurate TTS model training.

\subsection{Thai Text Dataset}\label{sec:Thai Text Dataset}
This section describes the data composition of our pure Thai Text Dataset, which includes a word corpus and a sentence corpus. Meticulously designed to ensure comprehensiveness and balance, the corpus serves as an optimal resource for a wide range of Thai language processing tasks while establishing a robust foundation for advanced linguistic research and computational applications in the field.


\begin{table*}[ht]
  \centering
  \small
  \begin{tabular}{c|l|c|l}
    \hline
    \textbf{Corpus} & \textbf{Data Source} & \textbf{Percentage} & \textbf{Description} \\
    \hline
    \multirow{4}{*}{Word} & PyThaiNlp  & 60\% & Lexicon from the PyThaiNlp tokenizer \\
    & Social Media and Online forums & 20\% & Twitter and Reddit public content\\ 
     & Official Corpora & 20\% & Open-source corpora from universities \\
    \cline{2-4}
    & \multicolumn{3}{|l}{\textit{Total: 100\% (100,000 words)}} \\
    \hline
    \multirow{6}{*}{Sentence} & News & 20\% & Curated news transcripts \\
    & Social Media & 10\% & Public posts from Thai social media platforms \\
    & E-books & 35\% & Text extracted from open-source e-books \\
    & Government Documents & 5\% & Official documents from government sources\\
    & Dictionaries & 30\% & Example sentences from dictionaries \\
    \cline{2-4}
    & \multicolumn{3}{|l}{\textit{Total: 100\% (1,000,000 sentences)}} \\
    \hline
  \end{tabular}
  \caption{Data composition of the Text Corpus.}
  \label{tab:text_data_composition}
\end{table*}

\noindent \textbf{Word Corpus}. The word corpus consists of the lexicon from the PyThaiNlp \cite{phatthiyaphaibun2023pythainlp} tokenizer (60,000 words) and the expanded vocabulary (40,000 words). The expanded vocabulary was manually selected by 20 native Thai speakers from social media, online forums and official corpora\footnote{Source:\url{https://www.arts.chula.ac.th/ling/tnc3/}} \footnote{Source:\url{https://aiforthai.in.th/corpus.php}} \footnote{Source:\url{https://belisan-volubilis.blogspot.com/}}, including technical terms, slang terms, neologisms and loanwords.

\noindent \textbf{Sentence Corpus}. The sentence corpus consists of data from news (20\%) \footnote{Source:\url{https://www.thairath.co.th}} \footnote{Source:\url{https://www.dailynews.co.th}} \footnote{Source:\url{https://news.sanook.com}} \footnote{Source:\url{https://www.thaipbs.or.th}} \footnote{Source:\url{https://www.manager.co.th}} \footnote{Source:\url{https://www.matichon.co.th}}, social media (10\%), e-books (35\%), government documents (5\%) \footnote{Source:\url{https://www.thaigov.go.th/main/contents}}, and dictionary example sentences (30\%)\footnote{Source:\url{http://www.thai-language.com/dict/}} \footnote{Source:\url{https://dict.longdo.com/}}. The dictionary data is based on Thai high-frequency word statistics, covering the top 50,000 most commonly used words. For each entry, 3–5 context sentences are crawled from multiple sources to match the word usage in different tenses and registers, ensuring semantic and syntactic diversity. During the preprocessing stage, a BERT-based cleaning model (based on WangchanBERTa \cite{lowphansirikul2021wangchanberta} pretraining) is employed to filter out duplicate, vulgar, or sensitive content. Sentences with high perplexity (PPL) are removed for semantic anomalies. Subsequently, the SentencePiece tokenization model \footnote{\url{https://github.com/google/sentencepiece}} is used to standardize sentence lengths to 10–25 words (long sentences are split, and short sentences are discarded). This process results in the construction of a high-quality corpus of one million sentences.

\subsection{Annotation Dataset}\label{sec:Annotation Dataset}
\textbf{Pause Annotation}: Of these, 2,000 sentences were manually annotated by 10 professional announcers according to Thai reading conventions, marking prosodic boundaries (short/long pauses, breathing points). Annotation consistency was verified using Kappa statistics ($\kappa = 0.82$). The remaining 3,000 sentences were segmented at the millisecond level using a high-precision voice activity detection (VAD) tool (WebRTC optimized version) on clean speech, supplemented by expert linguistic review to ensure alignment between automatic labeling and manual rules.

\noindent \textbf{Phoneme-Tone Annotation}: This task was completed by eight native Thai speakers trained in our annotation rules. 
\begin{figure}[htbp]
    \centering
    \includegraphics[width=0.5\textwidth]{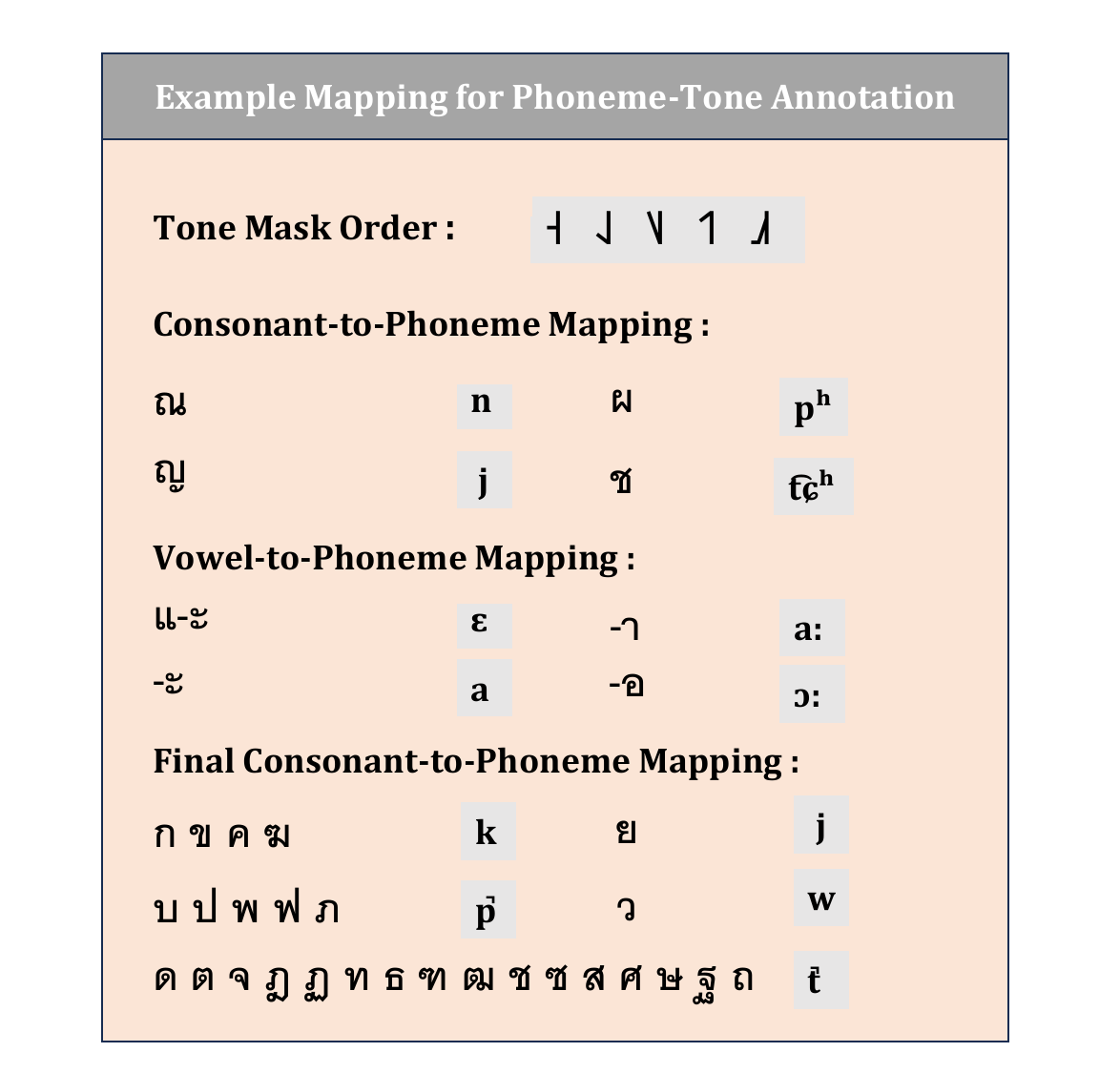}
    \label{fig:phoneme}
\end{figure}
After independent annotation of the full dataset, discrepancies (5.7\%) were submitted for arbitration by linguistic experts. The final annotation standards included: IPA Phonemes and Tone Symbols\footnote{Source:\url{https://thai-alphabet.com/}} \footnote{Source:\url{https://en.wikipedia.org/wiki/Help:IPA/Thai}}.

\end{appendices}
\end{CJK}
\end{document}